\documentclass[12pt]{elsarticle}

\usepackage{graphicx}
\usepackage{subfigure}
\usepackage{amsmath}
\usepackage{amssymb}
\usepackage{bm}

\journal{Phys. Lett. A}

\begin{document}

\title{Multiple phase transitions in the $XY$ model with nematic-like couplings}
\author{Milan \v{Z}ukovi\v{c}}
\ead{milan.zukovic@upjs.sk}
\address{%
Institute of Physics, Faculty of Science, P. J. \v{S}af\'arik University, \\
Park Angelinum 9, 040 01 Ko\v{s}ice, Slovakia
}%

\date{\today}

\begin{abstract}

Critical behavior of the two-dimensional generalized $XY$ model involving solely nematic-like terms of the second, third and fourth orders is studied by Monte Carlo method. We find that such a system can undergo three successive phase transitions. At higher temperatures there is a phase transition of the Berezinskii-Kosterlitz-Thouless type to the $q=4$ nematic-like phase, followed by two more transitions of the Ising type to the $q=2$ nematic-like and ferromagnetic phases, respectively. The $q$ nematic-like phases are characterized by spin alignments with angles $2k\pi/q$, where $k \leq q$ is an integer. The ferromagnetic phase appears at low temperatures even without the presence of magnetic interactions owing to a synergic effect of the nematic-like terms.

\end{abstract}


\begin{keyword}
Generalized $XY$ model, Nematic interactions, Berezinskii-Kosterlitz-Thouless phase, Ferromagnetic phase, Ising universality
\end{keyword}

\maketitle

\section{Introduction}

A generalized $XY$ model with the Hamiltonian ${\mathcal H_q}=-J_q\sum_{\langle i,j \rangle}\cos(q\phi_{i,j})$, where $J_q>0$ is restricted to nearest-neighbor pairs forming the angle $\phi_{i,j}=\phi_{i}-\phi_{j}$ and $q \in \mathbb{N}$, is known to show either magnetic or nematic-like quasi-long-range ordering (QLRO), for $q=1$ (the standard $XY$ model~\cite{bere71,kost73}) or $q>1$, respectively. Owing to the fact that the partition function of the latter can be mapped onto the former by the transformation $q\phi_{i} \to \phi_{i}$, the respective QLRO phases emerge at the same Berezinskii-Kosterlitz-Thouless (BKT) transition temperature, below which they can be characterized by bound pairs of either integer or fractional vortices and antivortices and the power-law decaying correlation function $g_q(r) = \langle \cos q(\phi_{0}-\phi_{r}) \rangle  \sim r^{-\eta_q}$~\cite{carme87}. While the correlation function $g_1$ is related to the ferromagnetic ordering in which spins have a common direction, $g_q$ with $q>1$ are related to only nematic-like axial alignments (of ``headless spins'') with angles $2k\pi/q$, where $k$ is an integer and $k \leq q$. Consequently, in the nematic-like phases there is no magnetic ordering and $g_1$ is expected to decay exponentially.  

Models that combine the $q=1$ and $q>1$ terms have been studied both out of theoretical curiosity (critical properties and universality) as well as in connection with various experimental realizations (e.g., liquid crystals~\cite{lee85,geng09}, superfluid A phase of $^3{\rm He}$~\cite{kors85}, high-temperature cuprate superconductors~\cite{hlub08}, DNA packing~\cite{grason08}, quasicondensation in atom-molecule, bosonic mixtures~\cite{bonnes12,bhas12,forg16}, and structural phases of cyanide polymers~\cite{cairns16,zuko16}). The most studied model, which included the $q=1$ and $q=2$ terms, has been shown~\cite{lee85,kors85,carp89,shi11,hubs13,qi13} to lead to the separation of the magnetic phase at lower and the nematic phase at higher temperature, for a sufficiently large nematic coupling. The high-temperature phase transition to the paramagnetic phase was determined to belong to the BKT universality class, while the magnetic-nematic phase transition had Ising character. 

Surprisingly, recent studies revealed that the model involving the $q=1$ and $q \geq 5$ terms displays a qualitatively different phase diagram featuring additional phases~\cite{pode11,cano16}. The latter appeared due to the competition between the respective couplings and the resulting phase transitions were determined to belong to different (Potts, Ising, or BKT) universality classes.

In our recent work we studied the generalized $XY$ model composed of solely nematic-like terms, with $q = 2$ and $q=3$~\cite{zuko18}. We found that, even though neither of the terms alone can induce magnetic ordering, their coexistence and competition can lead to a complex phase diagram including a magnetic phase at low temperatures. In particular, the ferromagnetic phase appeared wedged between the two nematic-like phases induced by the respective couplings. Thus, except for the muticritical point, at which all the phases meet, for any considered coupling parameters there was one transition from the paramagnetic phase to one of the two nematic-like phases followed by another one to the magnetic phase. While the phase transitions between the paramagnetic and nematic-like phases were of the BKT type, those between the magnetic and nematic-like phases were found to belong to the Ising and three-state Potts universality classes.

In the present study we extend our previous investigations by adding the fourth-order term, i.e., we consider the model that involves three purely nematic-like terms, with $q = 2,3$ and $4$, and explore possibility of any novel critical behavior. In particular, we demonstrate that for certain values of the parameters $J_2$,$J_3$, and $J_4$ even three different phase transitions are possible. Namely, the high-temperature order-disorder transition belonging to the BKT universality class is followed by two more phase transitions, as the system passes through two nematic-like phases to the magnetic phase at low temperatures, that both belong to the Ising universality class.

\section{Model and Method}

We consider the following Hamiltonian
\begin{equation}
\label{Hamiltonian2}
{\mathcal H}=-\sum_{q=2}^{4}J_q\sum_{\langle i,j \rangle}\cos(q\phi_{i,j}),
\end{equation}
where $\phi_{i,j}=\phi_{i}-\phi_{j}$ is an angle between the nearest-neighbor spins. In the following, the values of the respective exchange interactions are chosen to demonstrate the possibility of three successive phase transitions as $J_2=0.3$, $J_3=0.2$, and $J_4=0.9$. 

Monte Carlo (MC) simulations are performed on systems of a linear size $L$ with the periodic boundary conditions using the Metropolis algorithm. After discarding the first $10^5$ MC sweeps (MCS) the following $5 \times 10^5$ MCS are taken for calculation of thermal averages of various thermodynamic quantities in the equilibrium. Simulations are initialized in the paramagnetic phase by a random configuration. Then the temperature is gradually lowered by $\Delta T$ (measured in units $J/k_B$, where $k_B$ is the Boltzmann constant) and the simulation at $T-\Delta T$ starts from the last configuration obtained at $T$. Error bars are evaluated using the $\Gamma$-method~\cite{wolf04}.

Critical exponents at the phase transitions between the identified phases as well as within the QLRO phases are obtained by a finite-size scaling (FSS) analysis. The latter is performed using the reweighting techniques~\cite{ferr88,ferr89} from long time series ($10^7$ MCS after discarding $2 \times 10^6$ MCS for thermalization) obtained close to the transition point.  

We calculate the following quantities: the specific heat per spin $c$
\begin{equation}
c=\frac{\langle {\mathcal H}^{2} \rangle - \langle {\mathcal H} \rangle^{2}}{L^2T^{2}},
\label{c}
\end{equation}
the generalized magnetizations $m_q$, $q=1,2,3,4$,
\begin{equation}
m_q=\langle M_{q} \rangle/L^2=\left\langle\Big|\sum_{j}\exp(iq\phi_j)\Big|\right\rangle/L^2,
\label{oq}
\end{equation}
and the corresponding susceptibilities $\chi_{q}$
\begin{equation}
\label{chi_oq}\chi_{q} = \frac{\langle M_{q}^{2} \rangle - \langle M_{q} \rangle^{2}}{L^2T}.
\end{equation}
We note that $m_1$ is a standard magnetization and $m_q$, with $q>1$, represent $q$-nematic order parameters\footnote{We note that these quantities are not true order parameters as they all vanish in the thermodynamic limit.}. Furthermore, it is useful to calculate the following quantities:
\begin{equation}
\label{D2}D_{lq} = \frac{\partial}{\partial \beta}\ln\langle M_{q}^{l} \rangle = \frac{\langle M_{q}^{l} {\mathcal H}
\rangle}{\langle M_{q}^{l} \rangle}- \langle {\mathcal H} \rangle,\ l=1,2.
\end{equation}

At standard second-order phase transitions the order parameters~(\ref{oq}) and maxima of the quantities~(\ref{chi_oq}-\ref{D2}) scale with the system size as
\begin{equation}
\label{fss_chi}\chi_{q,max}(L) \propto L^{\gamma/\nu},
\end{equation}
\begin{equation}
\label{fss_D1}D_{lq,max}(L) \propto L^{1/\nu},\ l=1,2.
\end{equation}

In the QLRO phases the exponent $\eta$ of the algebraically decaying correlation function can be obtained from the FSS relation
\begin{equation}
\label{m_FSS}
m_q(L) \propto L^{-\eta/2}.
\end{equation}

\section{Results}

\begin{figure}[t!]
\centering
    \subfigure{\includegraphics[scale=0.42]{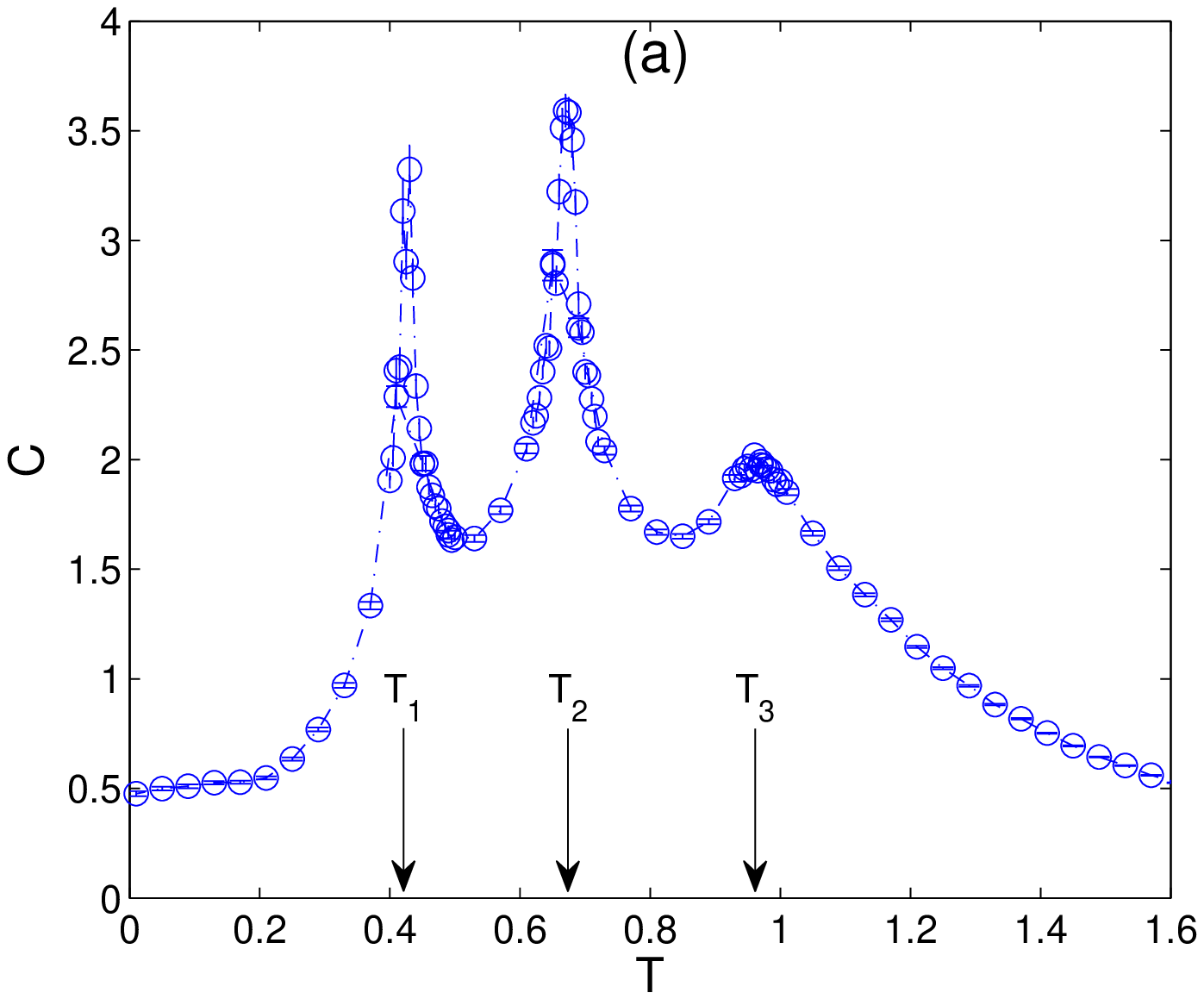}\label{fig:c-T}}
		\subfigure{\includegraphics[scale=0.42]{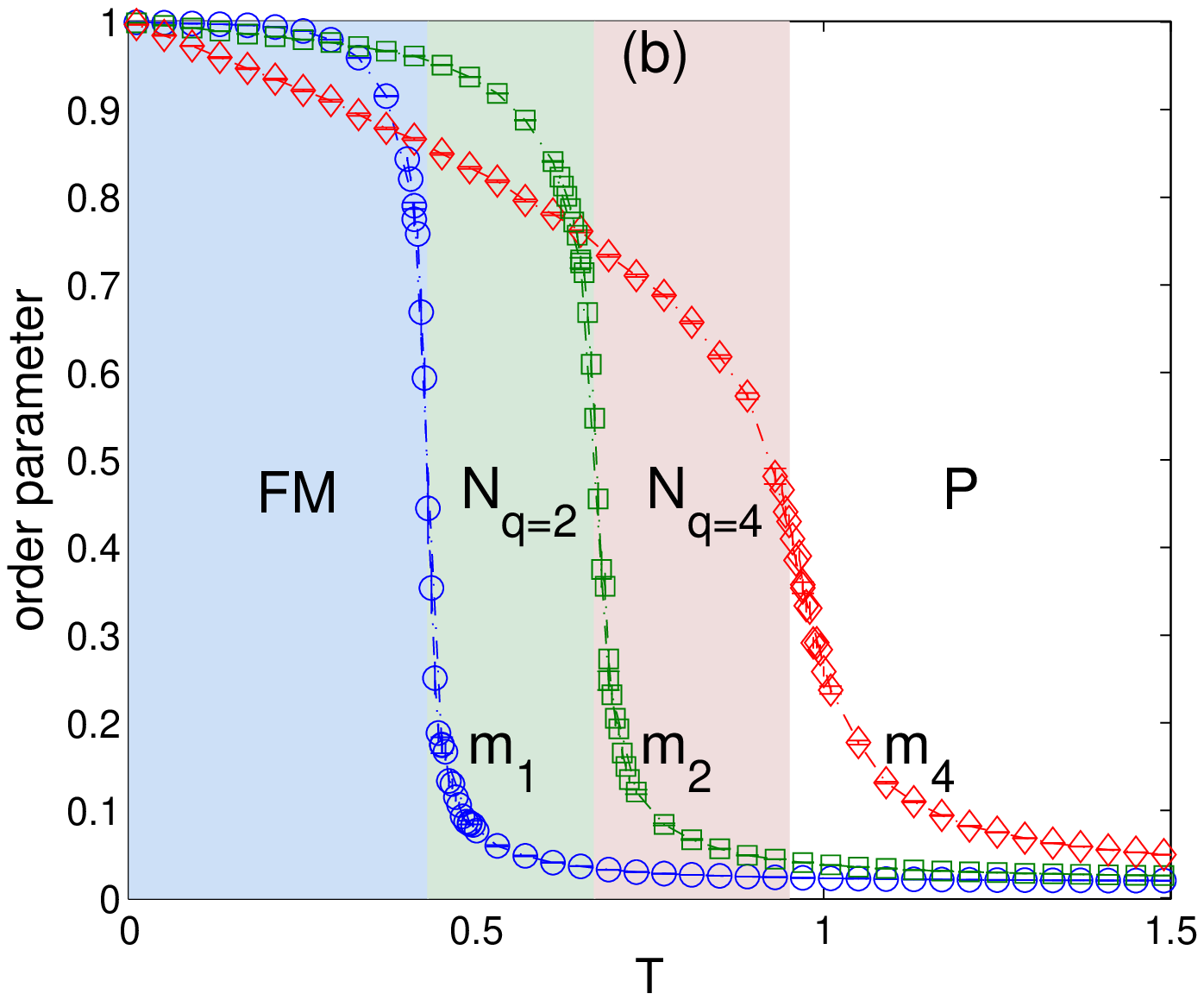}\label{fig:o-T}}
\caption{(Color online) Temperature variations of (a) the specific heat $c$ and (b) the order parameters $m_q$, for $L=48$. FM, $N_{q=2}$, $N_{q=4}$ and $P$ stand for the ferromagnetic, $q=2$ nematic, $q=4$ nematic and paramagnetic phases, respectively.}\label{fig:x-T}
\end{figure}

Phase transitions can be detected and roughly localized from anomalies (peaks) of various quantities, such as the response functions (\ref{c}) and (\ref{chi_oq}). In Fig.~\ref{fig:c-T} we show the temperature dependence of the specific heat, for $L=48$. One can clearly distinguish peaks at  three different temperatures $T_1<T_2<T_3$. While the high-temperature peak is rounded and less pronounced, the other two peaks at lower temperatures are sharp and much higher. The difference suggests different characters of the respective phase transitions, which will be analyzed in more detail below.

To determine the nature of the respective phases separated by the transition points, we plot in Fig.~\ref{fig:o-T} different order parameters introduced in Eq. (\ref{oq}). The plot shows that $T_1$, $T_2$ and $T_3$ are the temperatures at which the generalized magnetizations $m_1$, $m_2$ and $m_4$, respectively, vanish\footnote{For clarity, $m_3$ is not shown but it vanishes together with $m_1$.}. Therefore, the system shows three QLRO phases: the ferromagnetic (FM) at low temperatures $T<T_1$, the $q=2$ nematic ($N_{q=2}$) at intermediate temperatures $T_1<T<T_2$ and the $q=4$ nematic ($N_{q=4}$) at higher temperatures $T_2<T<T_3$. It is worth noting that the FM phase appears in spite of the absence of magnetic interactions. Its emergence can be attributed to a synergic effect of the present nematic-like interactions, as also observed in the model with only $q = 2$ and $q=3$ terms~\cite{zuko18}.

\begin{figure}[t]
\centering
    \subfigure{\includegraphics[scale=0.42]{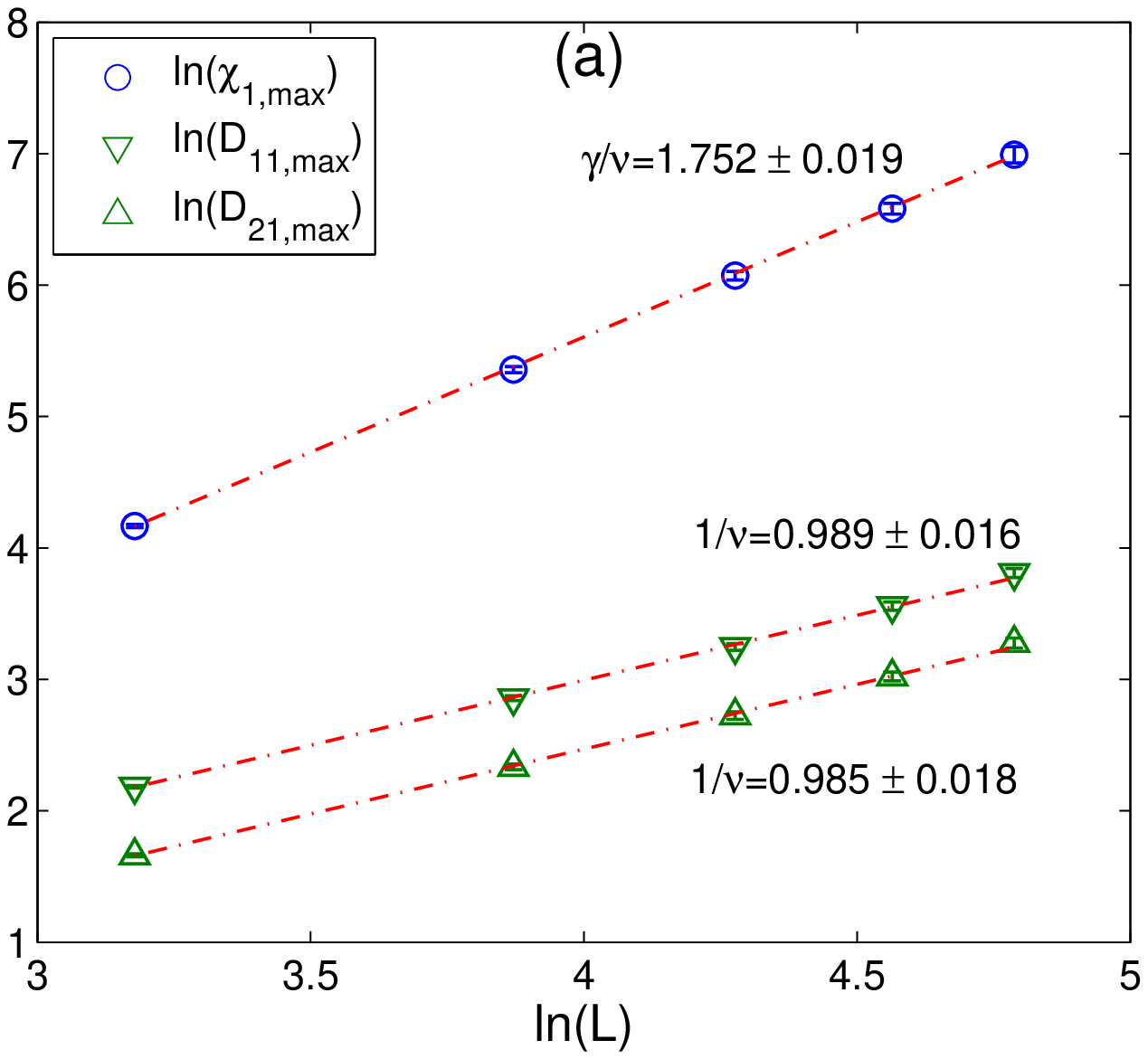}\label{fig:fss_T_0_42_m}}
		\subfigure{\includegraphics[scale=0.42]{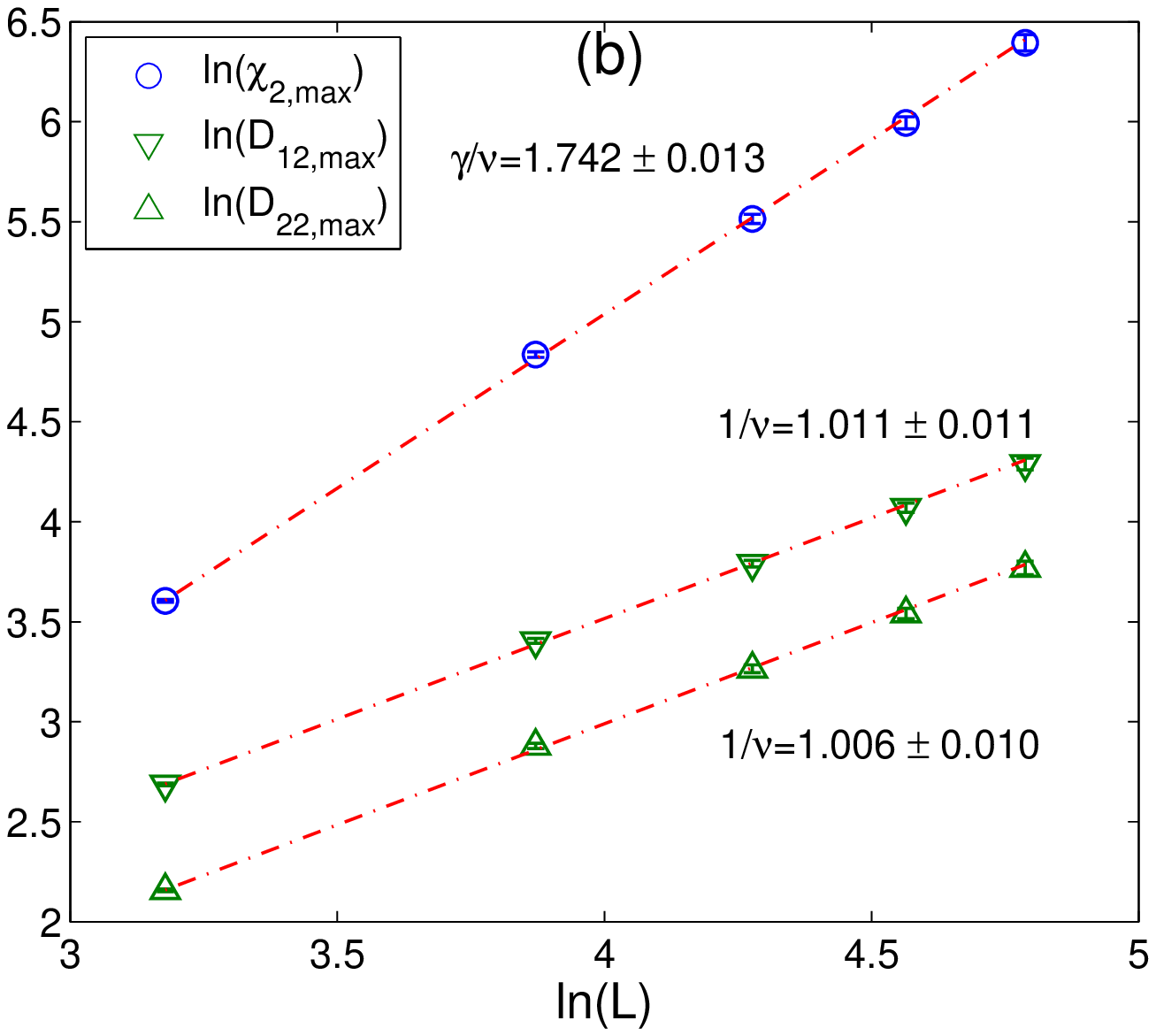}\label{fig:fss_T_0_67_n2}}
\caption{(Color online) Critical exponents ratios at (a) the $FM-N_{q=2}$ and (b) $N_{q=2}-N_{q=4}$ phase transitions.}
\label{fig:fss-I}
\end{figure}

In the next step we perform a FSS analysis in effort to determine the character of the respective phase transitions (universality class). In particular, we calculate the critical exponents from the scaling relations (\ref{fss_chi}) and (\ref{fss_D1}). The results obtained at the $FM-N_{q=2}$ and $N_{q=2}-N_{q=4}$ transition points are presented in Fig.~\ref{fig:fss-I}. In both cases the critical exponent ratios coincide within statistical errors with the Ising universality class values $1/\nu_{\rm I}=1$ and $\gamma_{\rm I}/\nu_{\rm I}=7/4$.

\begin{figure}[t]
\centering
\includegraphics[scale=0.5]{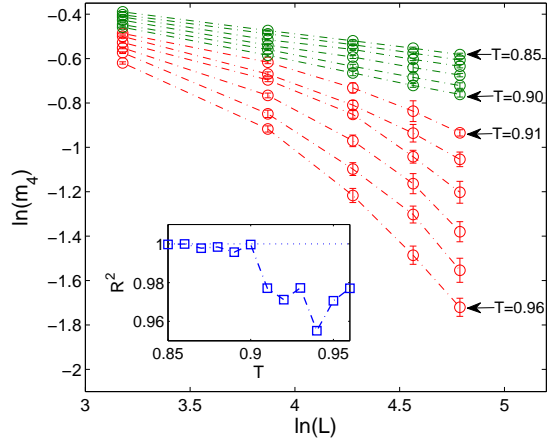}
\caption{(Color online) Log-log plot of the generalized magnetization $m_4$ against the lattice size, for different temperatures. The inset shows the adjusted coefficient of determination $R^2$.}
\label{fig:fss_m4_BKT}
\end{figure}

Finally, let us focus on the $N_{q=4}-P$ transition. It is well known that in the case of the BKT transition the specific heat peak position overestimates the true value of the transition temperature. Therefore, in order to estimate it with higher precision we apply an approach based on the study of the correlation function behavior. In particular, we perform a FSS analysis of the parameter $m_4$, which is expected to vanish in the algebraic phase according to the scaling relation~(\ref{m_FSS}), for various temperatures. Then, the transition point is determined as a point at which the algebraic dependence changes to the exponential one. In Fig.~\ref{fig:fss_m4_BKT}, it is apparent that the curves corresponding to the lower temperatures $(T \leq 0.9)$ show linear dependence, as expected within the algebraic BKT phase. On the other hand, the curves at higher temperatures start showing a downturn, as expected in the paramagnetic phase with the exponentially decaying correlation function. The transition is also reflected in the sudden drop of the adjusted coefficient of determination (a measure of goodness of the linear fit) below $R^2 = 1$, as shown in the inset. Thus the $N_{q=4}-P$ transition temperature can be estimated as $0.90 < T_{\rm BKT} < 0.91$.

\section{Summary and discussion}

In summary, we studied the generalized $XY$ model, comprising three purely nematic-like terms, and found that there exists a region in the exchange parameter space in which the system can display three different phase transitions. The high-temperature order-disorder transition, that belongs to the BKT universality class, is followed by two more phase transitions, that belong to the Ising universality class. The respective phases have the character of the $q=4$ nematic, $q=2$ nematic and ferromagnetic quasi-long-range order. 

The aim of the present Letter was to demonstrate the existence of the region in the $J_2-J_3-J_4$ parameter space showing multiple phase transitions. However, a more systematic study would be desirable to establish the phase diagram in the entire parameter space in order to determine its extent. Nevertheless, such investigations by MC simulations would require immense computational effort and it is out of the present scope. Furthermore, in the light of the recently found new phases in the model involving only two terms with $q=1$ and $q \geq 5$~\cite{pode11,cano16}, it would be interesting to explore the possibility of the existence of additional phases in the generalized $XY$ model involving multiple terms with the $q$-values ranging between $q_{min} = 1$ and $q_{max} \geq 5$.

\section*{Acknowledgments}
This work was supported by the Scientific Grant Agency of Ministry of Education of Slovak Republic (Grant No. 1/0331/15) and the scientific grants of Slovak Research
and Development Agency provided under contracts No.~APVV-16-0186 and No.~APVV-14-0073.

\end{document}